\documentclass[aip,apl,twocolumn,numerical]{revtex4}

\usepackage{times}
\usepackage{graphicx,epstopdf}
\usepackage{pslatex}
\usepackage{amssymb}
\usepackage{amsmath}
\usepackage[caption=false]{subfig}

\begin{document}

\title{Manipulation of Giant Faraday Rotation in Graphene Metasurfaces}

\author{Arya Fallahi} \affiliation{DESY-Center for Free-Electron Laser Science, Hamburg University, Notkestrasse 85, D-22607 Hamburg, Germany}
\author{Julien Perruisseau-Carrier} \affiliation{Ecole Polytechnique F\'{e}d\'{e}rale de Lausanne (EPFL), CH-1015 Lausanne, Switzerland}

\date{\today}

\begin{abstract}

Faraday rotation is a fundamental magneto-optical phenomenon used in various optical control and magnetic field sensing techniques.
Recently, it was shown that a giant Faraday rotation can be achieved in the low-THz regime by a single monoatomic graphene layer. 
Here, we demonstrate that this exceptional property can be manipulated through adequate nano-patterning, notably achieving giant rotation up to 6THz with features no smaller than 100nm. 
The effect of the periodic patterning on the Faraday rotation is predicted by a simple physical model, which is then verified and refined through accurate full-wave simulations.

\end{abstract}

\pacs{33.57.+c and 81.05.Xj}
\maketitle

Thanks mainly to its unconventional band structure and two-dimensional nature \cite{theRiseOfGraphene,grapheneStatusAndProspects}, graphene exhibits many exceptional electrical and optical properties such as high electron mobility \cite{masslessDiracFermions1,masslessDiracFermions2}, quantum Hall effect \cite{quantumHallEffect}, highly confined plasmonic propagation \cite{graphenePlasmonics}, large nonlinear Kerr effect \cite{grapheneKerrEffect} and giant Faraday rotation \cite{GiantFaradayRotation}. 
Moreover, these properties can be dynamically tuned through electrostatic and magnetostatic bias, which is another major feature of graphene. 
In particular, Faraday rotation, which is the focus of this work, is used in various applications in optical telecommunications and laser technology (modulators, optical isolators and circulators) \cite{faradayEffect}, as well as for advanced magnetic field sensing and even in spintronics \cite{spintronics}. 

Nevertheless, despite the tuning possibility, the general behavior of graphene devices is determined by the intrinsic characteristics of the material and can hardly be tailored for a specific application or wavelength of interest. 
In the case of the giant Faraday rotation, a $7\,\mbox{T}$ magnetic bias allows the dynamic differential rotation of a plane wave polarization up to about 6 degrees. 
However, the rotation is maximum only at low frequencies and decreases for smaller wavelengths, notably reducing very rapidly above 2THz \cite{GiantFaradayRotation}. 
In \cite{grapheneMicrowave}, a scheme based on electric biasing of graphene and consequently increasing the chemical potential is proposed to tailor the Faraday rotation property of graphene for microwave applications.

Following the advent of the metamaterial concept, it was shown that patterning bulk materials to control their electromagnetic properties enables many fantastic applications such as negative refraction, photonic bandgap and perfect lenses \cite{bookmetamaterial1,bookmetamaterial2}. 
This idea was then applied to graphene in \cite{patternedGrapheneNature} for the control of the transmission magnitude, and other functionalities such as absorbers and cloaks were also recently suggested in \cite{grapheneCloaking}. 
Micro-patterning of graphene monolayers can be achieved for instance using electron beam lithography (EBL) and oxygen plasma etching \cite{patternedGrapheneNature}, which confirms the applicability of graphene micro-patterning for optical devices. 
Here, we theoretically demonstrate that one of the most amazing properties of graphene, namely, the single-atomic layer giant Faraday rotation, can be efficiently manipulated through nano-patterning. 
It is notably predicted that giant Faraday rotation can be achieved at much higher frequencies than using a uniform unpatterned layer. 
Our approach consists in first comparing our numerical results with formerly measured results for uniform layers, and then predicting the effect of nano-patterning on the Faraday response. 
The results are also shown to be in good agreement with a proposed simple physical circuit model.

\begin{figure}[!h]
\centering
\includegraphics[width=3.2in]{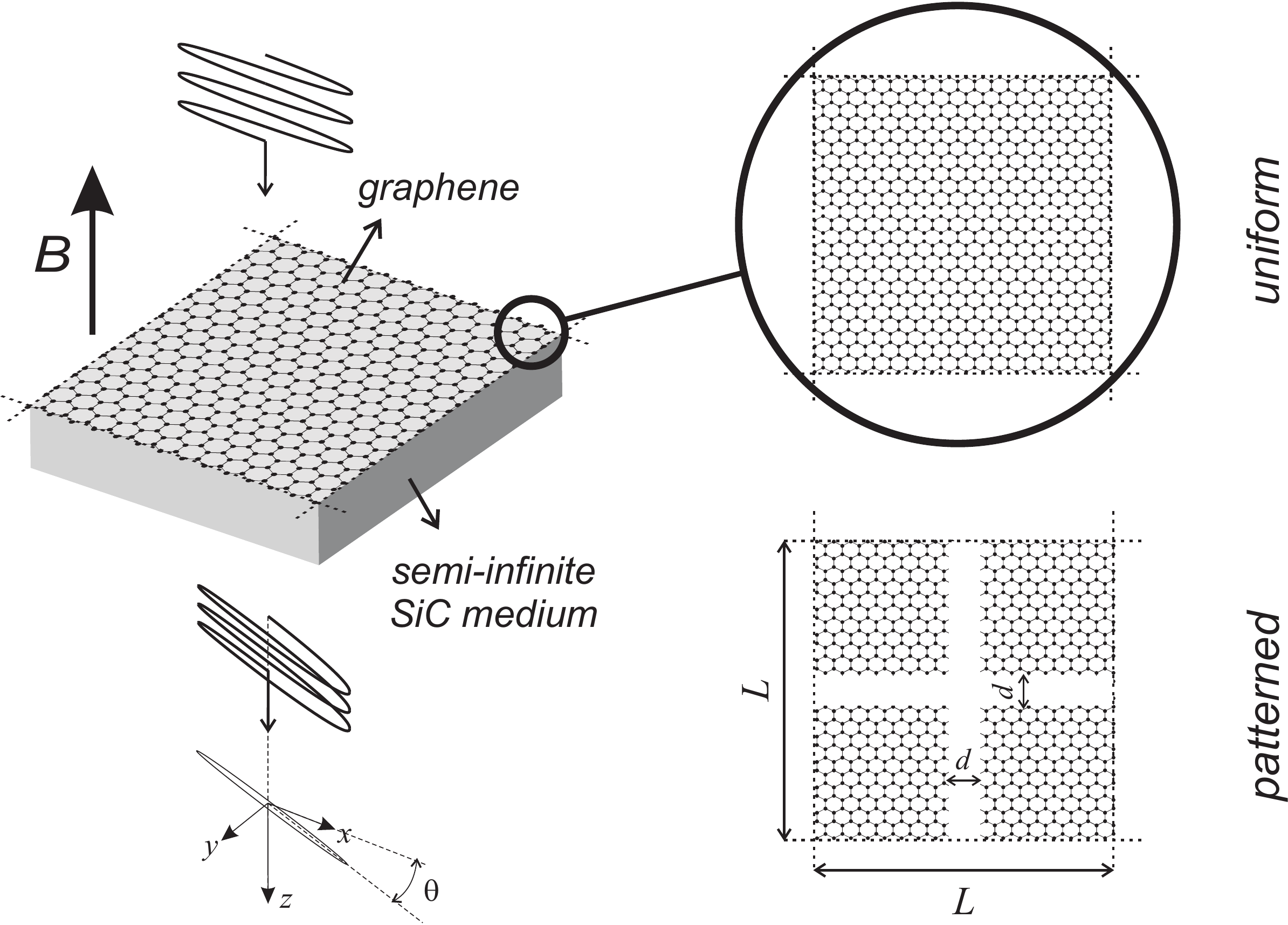}
\caption{Illustration of Faraday rotation and the unit cells of both uniform and patterned graphene layers}
\label{Fig1}
\end{figure}

\begin{figure}[!h]
\centering
\includegraphics[width=3.375in]{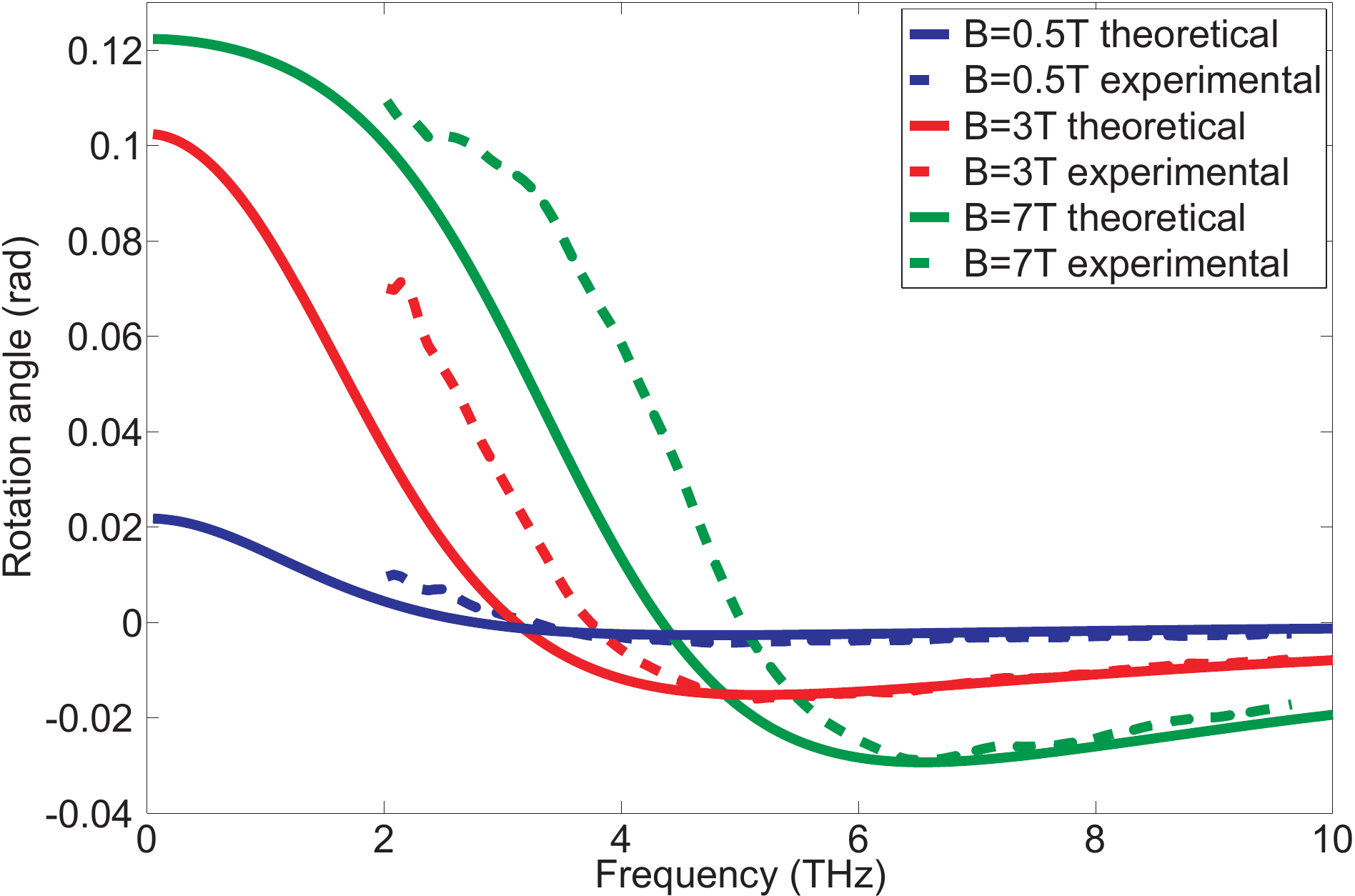}
\caption{Rotation angle of the transmitted plane wave versus frequency for a uniform graphene monolayer residing on SiC substrate under different magnetostatic biases, experimental (see Ref. \cite{GiantFaradayRotation}) and theoretical}
\label{Fig2}
\end{figure}

Consider first a uniform graphene layer on an infinitely thick SiC substrate (Fig.\,\ref{Fig1}) with Dirac voltage $V_d=0.34\,\mbox{eV}$, temperature $T=5\,\mbox{K}$ and phenomenological scattering rate $\Gamma = 10\,\mbox{meV}$ \cite{GiantFaradayRotation}. 
The frequency-dependent Faraday rotation numerically computed using Kubo-formalism \cite{grapheneConductivityModel}, as well as the corresponding measurements \cite{GiantFaradayRotation}, are shown in Fig.\,\ref{Fig2} for different magnetic bias fields. 
The results are in fair agreement, the discrepancies being due to defects in the fabricated graphene layer \cite{intrinsicPlasmons}. 

\begin{figure}[h]
\centering
\includegraphics[width=2.5in]{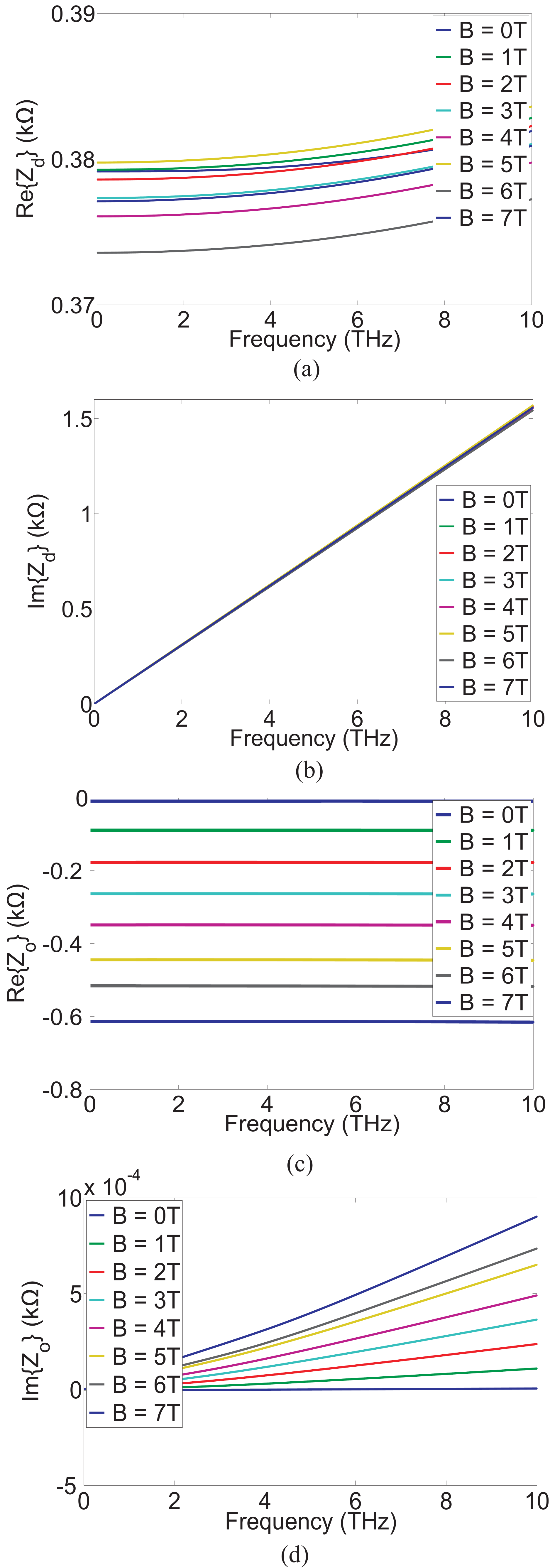}
\caption{(a) real part of $Z_d$, (b) imaginary part of $Z_d$, (c) real part of $Z_o$ and (d) imaginary part of $Z_o$ in the anisotropic impedance matrix of a magnetically biased graphene monolayer}
\label{Fig3}
\end{figure}
The Faraday rotation can be well approximated as
\begin{equation}
\label{FaradayRotationFrequency}
\theta (\omega,B) \simeq Z_0 f(\omega) \Re \{ \sigma_{o} (\omega,B) \},
\end{equation}
where $Z_0$ is the impedance of vacuum, $f(\omega)$ depends on the substrate properties and $\sigma_{o} (\omega,B)$  is the off-diagonal term in the conductivity tensor of graphene ($\mbox{\boldmath$\sigma$} = [ \sigma_d \,\,\, \sigma_o ; -\sigma_o \,\,\, \sigma_d ]$), itself calculated using Kubo's formalism \cite{GiantFaradayRotation}. 
Graphene anisotropic conductivity closely follows a Drude conductivity model at THz. 
Hence, both diagonal and off-diagonal terms of the corresponding surface impedance can be accurately expressed as a superposition of resistive and inductive components.  
Indeed, by computing the impedance matrix
\begin{equation}
\label{grapheneImpedance}
Z = \mbox{\boldmath$\sigma$}^{-1} = \left( \begin{array}{cc} Z_d & -Z_o \\ Z_o & Z_d \end{array} \right),
\end{equation}
one obtains the impedance values shown in Fig.\,\ref{Fig3}. 
The plots reveal that both the diagonal and off-diagonal terms are very-well approximated by an equivalent surface R-L impedance model, i.e. $Z_d=R_d+j\omega L_d$ and $Z_o=R_o+j\omega L_o$ with $R_d \simeq 380\,\mathrm{\Omega}/\square$, $L_d \simeq 25.5\,\mbox{pH}/\square$, $R_o \simeq - 87.1 \cdot B\,\mathrm{\Omega}/\square$, and $L_o \simeq 2.05 \times 10^{-3} \cdot B\,\mbox{pH}/\square$. 
It is observed that the off-diagonal terms depend on the magnitude of the bias magnetic field $B$ (in Tesla), which is obviously the origin of the possibility for dynamically controlling Faraday Rotation through magnetic biasing. 

Following this circuit model representation, we inferred that artificially creating some electrical energy reactive field storage along the graphene layer would allow further control of $\sigma_o (\omega,B)$. 
From a practical perspective and inspired by the metamaterial concept, the simplest way to achieve the desired capacitive effect is to periodically pattern out narrow strips of graphene. 
Indeed, consider now graphene periodically patterned as in Fig.\,\ref{Fig1}. 
The stored energy associated with the electric field in the gaps translates in an additional isotropic capacitive effect. 
In addition, the overall relative surface of graphene is reduced and thus the inherent resistive and inductive graphene impedances scale accordingly. 
Therefore, the total effective impedance can be written as follows:
\begin{equation}
\label{totalImpedance}
Z_t = \left(\frac{L}{L-d}\right)^2 \left( \begin{array}{cc} Z_d & -Z_o \\ Z_o & Z_d \end{array} \right) + \left( \begin{array}{cc} \frac{1}{j\omega C} & 0 \\ 0 & \frac{1}{j\omega C} \end{array} \right),
\end{equation}
Now, the effective capacitance associated with the periodic pattern considered here can be estimated as \cite{periodicGridCapacitance} 
\begin{equation}
\label{capacitance}
C = \epsilon_0 L \frac{\epsilon_r+1}{\pi} \ln (\csc \frac{\pi d}{2L} ),
\end{equation}
where $\epsilon_r$ is the dielectric permittivity of the substrate. 
Although this equation was derived for perfect conductors, the high electron mobility in graphene justifies this assumption for evaluating the distributed capacitive effect here. 

\begin{figure}[!h]
\centering
\includegraphics[width=3.375in]{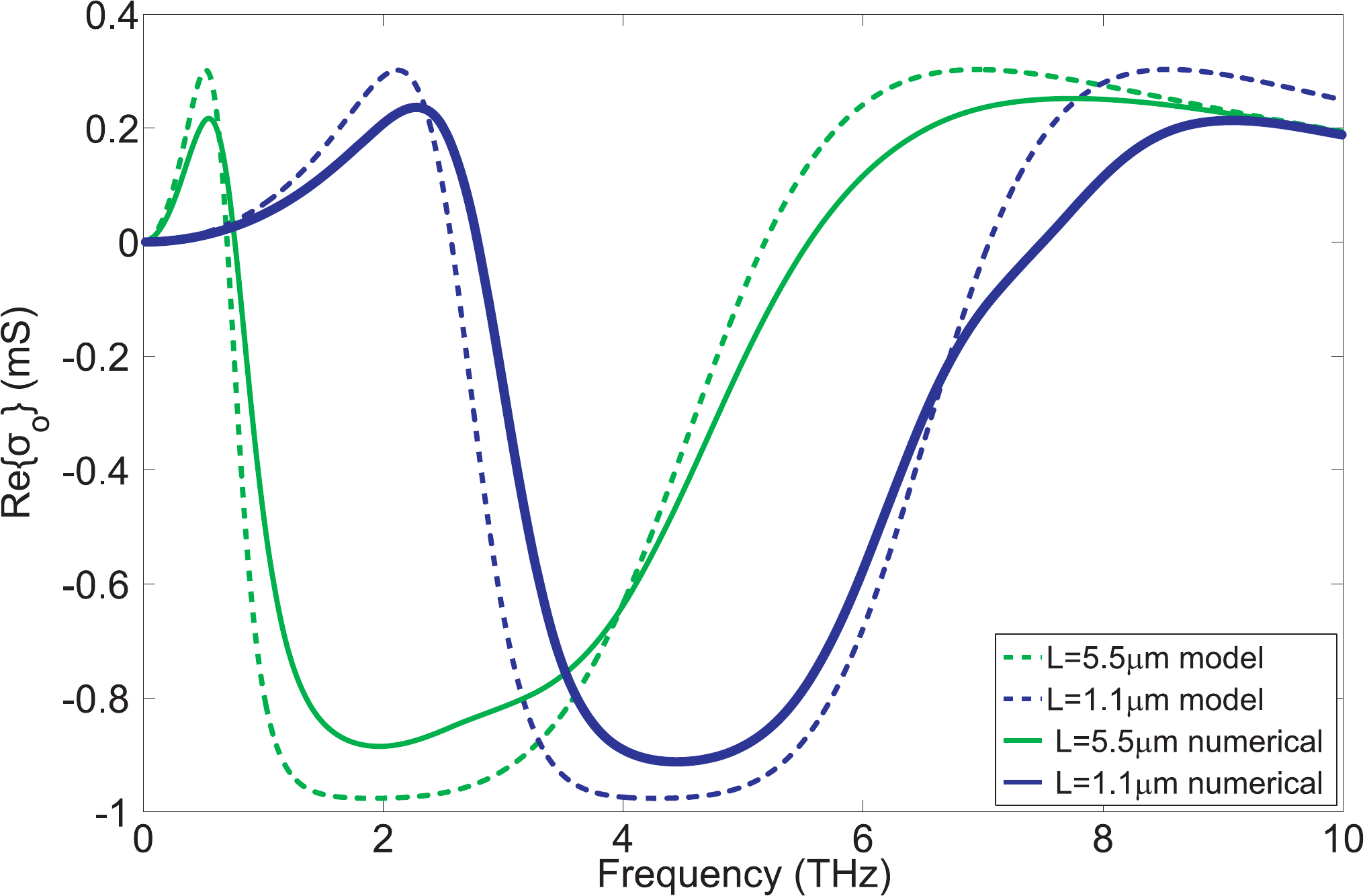}
\caption{Real part of the effective off-diagonal conductivity, proposed physical model versus accurate full-wave approach}
\label{Fig4}
\end{figure}

The proposed model was used to compute the effective off-diagonal conductivity of a patterned graphene monolayer with different lattice constants and $B=7\,\mbox{T}$.
The same parameter was also computed using an in-house full-wave simulator based on the periodic Method of Moment \cite{myThesis}, which is able to handle graphene anisotropic conductivity as well as Floquet's periodic boundaries to rigorously account for the coupling between unit cells. 
Note that the results of the code exactly fit the analytical formula available in the case of uniform graphene layer (Fig.\,\ref{Fig2}). 
Fig.\,\ref{Fig4} shows the comparison of both modeling methods for two representative cases.
The good agreement demonstrates the validity of the above physical interpretation, as well as the ability of the simple proposed model to provide good prediction of the frequency and bias dependent Faraday rotation. 

\begin{figure}[!h]
\centering
\includegraphics[width=3.375in]{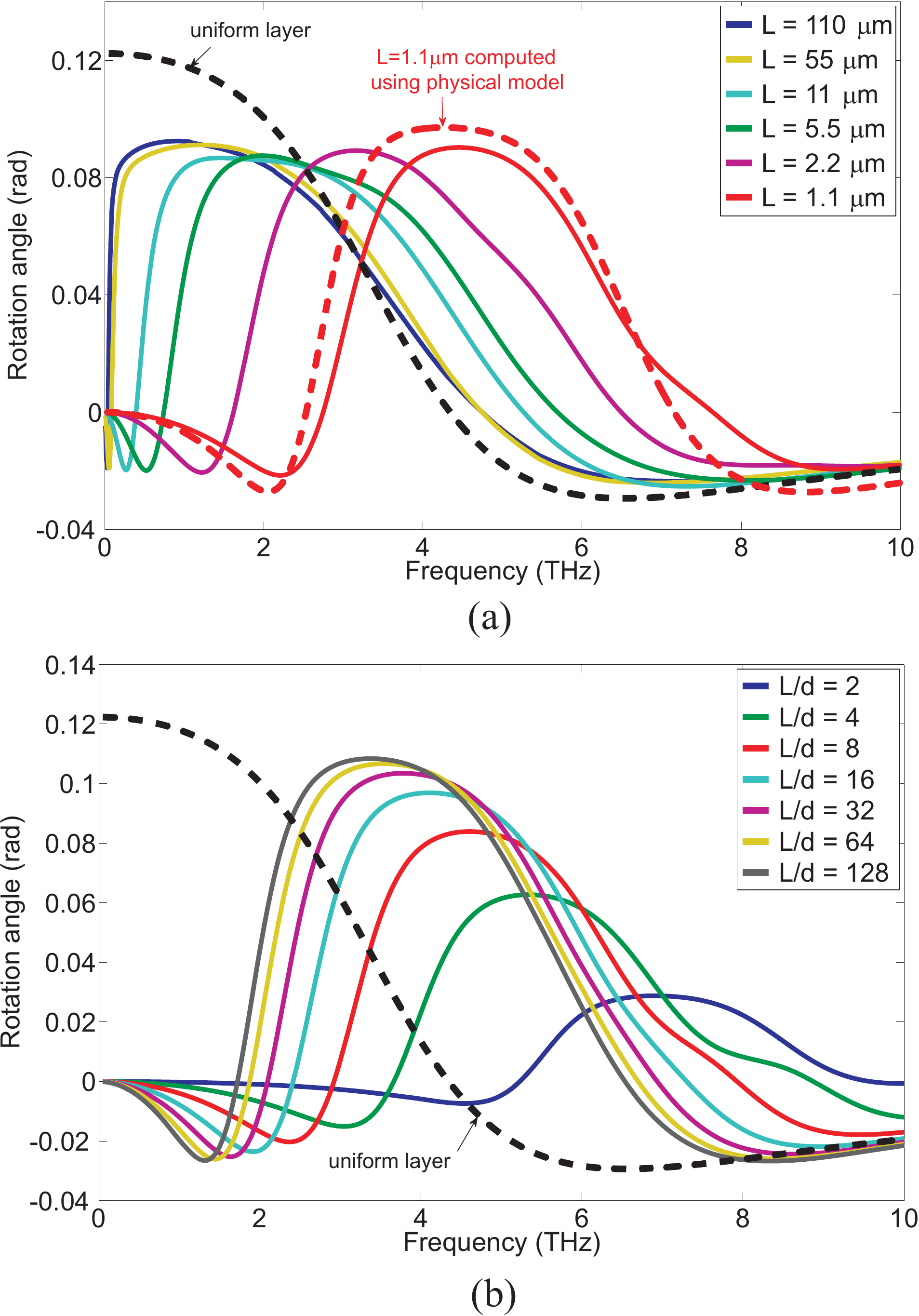}
\caption{(a) Rotation angle of patterned graphene for different lattice constants and $7\,\mbox{T}$ magnetic bias, with gap length fixed to one-eleventh of the lattice constant, i.e. $d=L/11$ and (b) Rotation angle of patterned graphene for different gap widths and $7\,\mbox{T}$ magnetic bias with the lattice constant fixed to $L=1.1\mu m$}
\label{Fig5}
\end{figure}

The corresponding Faraday rotations are shown in Fig.\,\ref{Fig5}. 
The graphs again include the above physically-based approximate model compared with the detailed full-wave solution. 
These results clearly validate the physical interpretation and model provided, as well as the possibility for manipulation of the Faraday rotation in terms of frequency and magnitude. 
The highest frequency of giant Faraday rotation in Fig.\,\ref{Fig5} is between 4THz and 6THz, a region where that of the uniform graphene is much weaker and strongly frequency-dependent. 
In fact, this case corresponds to a gap width $d=100\,\mbox{nm}$ that is easily achievable via e-beam lithography, thereby demonstrating the practical feasibility of the concept to such (and even smaller) wavelengths. 
Comparison between Fig.\,\ref{Fig5} also reveals that the amount of maximum giant Faraday rotation mostly depends on the overall relative graphene area in the patterned surface (this value is constant in Fig.\,\ref{Fig5}a),  while the frequency at which the maximum rotation is obtained is controlled by the absolute value of $L$ and $d$. 

The reported phenomenon provides a prominent degree of freedom for applications of giant Faraday rotation and the associated magneto-optical Kerr effects. 
Through the patterning of a graphene layer, a dynamically-controllable giant Faraday rotation can be obtained at a frequency determined a priori by the patterning dimensions. 
This technique can potentially lead to a variety of applications such as ultrathin circulators and isolators for both infrared and THz, miniaturized Faraday rotators for amplitude modulation, or integrated magnetic field sensors. 
From a more general perspective, these results motivate the further study of graphene patterning for a much better and versatile control of numerous exceptional electromagnetic properties.

\section{Acknowledgement}

The authors would like to thank Alexey B. Kuzmenko and the other authors of \cite{GiantFaradayRotation} for kindly providing the measurements data of giant Faraday rotation in Fig.\,\ref{Fig2}.
This work was supported by the Swiss National Science Foundation (SNSF) under grant n$^{\circ}$133583.

\end{document}